\documentclass[english,12pt]{article}
\usepackage{color}
\usepackage{amssymb}
\usepackage{amsmath}
\usepackage{babel}
\usepackage{pifont}
\usepackage
[dvips,
colorlinks=true,
linkcolor=webgreen,
filecolor=webbrown,
citecolor=webgreen,
bookmarksopen=false,
]{hyperref}
\definecolor{webgreen}{rgb}{0,.5,0}
\definecolor{webbrown}{rgb}{.6,0,0}
\usepackage[T1]{fontenc}
\usepackage[cp1250]{inputenc}
\usepackage{array}
\usepackage{amssymb}
\date{}
\definecolor{arcolor}{cmyk}{0.05,0.95,0.9,0.1}
\title{Quantum computer: an appliance for playing  market games}
\author{Edward W. Piotrowski\\ Institute of Theoretical Physics,
University of Bia\l ystok,\\ Lipowa 41, Pl 15424 Bia\l ystok,
Poland\\ e-mail: \href{mailto:ep@alpha.uwb.edu.pl}{ep@alpha.uwb.edu.pl}\\
 Jan S\l adkowski\\ Institute of Physics, University of Silesia, \\ Uniwersytecka
4, Pl 40007 Katowice, Poland \\ e-mail:
\href{mailto:sladk@us.edu.pl}{sladk@us.edu.pl} }
\usepackage[dvips]{graphicx}
\usepackage{pstricks}
\usepackage{pst-node}
\begin{document}
\maketitle
\def\Z{{\bf Z\!\!Z}}
\def\R{{\bf I\!R}}
\def\N{{\bf I\!N}}
\def\C{{\bf I\!\!\!\! C}}
\def\meter{\mbox{$\frown\hspace{-.9em}{\lower-.4ex\hbox{$_\nearrow$}}$}}
\begin{abstract}
Recent development in quantum computation and quantum information
theory  allows to extend the scope of game theory for the quantum
world. The authors have recently proposed a quantum description of
financial market in terms of quantum game theory. The paper
contain an analysis of such markets that shows that there would be
advantage in using quantum computers and quantum strategies.
\end{abstract}


 \vspace{5mm}

\section{Motivation}
Most of reasonable people find their best ways of doing something
by analysis of previously met situations. This rule usually
results in moves that are profitable or beneficial in a more or
less general sense. Sciences and martial arts have worked out an
additional nonconservative scheme that consist in analysis of the
events in their reversed historical perspective (one revises the
axioms or decisions bearing  their consequences in mind). Shall we
believe that traders active on the future markets who would have
sophisticated achievement of quantum technology at their disposal
would refrain from using them \cite{1}? Below we will discuss some
selected aspects of quantum markets \cite{2}-\cite{4} that may
form the basis for extraordinary profits. By changing some of our
habits we might be able to gain unique profits resulting from
discoveries made by researchers in fundamental natural phenomena
that suggest more effective ways of playing games \cite{5}. These
"new games" cannot by itself create extraordinary profits or
multiplication of goods but the dynamism of transaction they may
cause would result in more effective markets and capital flow into
hands of the most efficient traders. Sophisticated technologies
that are not yet available are not necessary to put such a market
in motion. Simulation of such markets can be performed in an
analogous way to precision physical measurements during which
classical apparatuses are used to explore  quantum phenomena.
People seeking after excitement would certainly not miss the
opportunity to perfect their skills at using "quantum strategies".
To this end an automatic game "Quantum Market" will be sufficient.
And such a device can be built up  due to the recent advances in
technology.

\section{Interference of traders decisions}
Let us consider  trading in some commodity $\mathfrak{G}$
according to as simple as possible but quantum decision rules
\cite{1,2,5}. The analyzed below game is certainly feasible with
contemporary physical instruments but we refrain from stating what
 the necessary technical requirements for implementation of
such games are. Let the states $|0\rangle$ and $|\text{I}\rangle$
denote strategies that the trader selling $\mathfrak{G}$ accepts a
low and a high price, respectively \cite{2}. The family
$\{|z\rangle\}$,$z\negthinspace\in\negthinspace\overline{\mathbb{C}}$
of complex vectors (states)
$|z\rangle:=|0\rangle+z\,|\text{I}\rangle$
($|\negthinspace\pm\negthinspace\infty\rangle:=|\text{I}\rangle$)
represents all trader strategies in the linear hull  spread by the
vectors $|0\rangle$ and $|\text{I}\rangle$. The coordinates of the
vector $|z\rangle$ (we will often call it a qubit to follow the
quantum information theory convention) in the basis
$(|0\rangle,|\text{I}\rangle)$ give after normalization
probability amplitudes of the corresponding traders decisions. It
is convenient to identify the strategies $|z\rangle$ with points
of the two dimensional sphere
$S_2\negthinspace\simeq\negthinspace\overline{\mathbb{C}}$. In
that case we can use the geographic coordinates $(\varphi,\theta)$
where $\varphi:=\arg z$ and $\theta:=\arctan|z|$. According to the
quantum model of market \cite{2,6,7} the strategy $|z\rangle$
expressed in the basis $(|0'\rangle,|\text{I}'\rangle)$ that
consists of Fourier transforms of the vectors
$(|0\rangle,|\text{I}\rangle)$\footnote{The bases
$(|0'\rangle,|\text{I}'\rangle)$ and
$(|0\rangle,|\text{I}\rangle)$ exemplify the notion of conjugate
bases that correspond to "extreme"  nonmeasurable simultaneously
observables in a finite dimensional Hilbert space, cf \cite{8}.}
\begin{equation}
|z\rangle=|0\rangle+z\,|\text{I}\rangle=|0'\rangle+\tfrac{1-z}{1+z}\,|\text{I}'\rangle
\label{herkkk}
\end{equation}
describes the trader's decisions concerning buying the commodity
$\mathfrak{G}$. In that sense the demand  aspects of trader's
behavior are Fourier representation of her "supplying
preferences". The involutive homography ${\mathcal
F}\negthinspace:\negthinspace|n\rangle\rightarrow
|n'\rangle=\frac{1}{\sqrt{2}}\sum_{m=0}^\text{I}(-1)^{\langle
n|m\rangle}|m\rangle $, $n\negthinspace=\negthinspace0,\text{I}$
that describes the Fourier transform in two dimensions has the
form of Hadamard matrix in both bases
$(|0\rangle,|\text{I}\rangle)$ and
$(|0'\rangle,|\text{I}'\rangle)$:
 $\frac{1}{\sqrt{2}}\begin{pmatrix}
  1 & \phantom{-}1 \\
  1 & -1
\end{pmatrix}$. The vectors $|0'\rangle$ and $|\text{I}'\rangle$ correspond to the
acceptance of a high  and a low price of the commodity
$\mathfrak{G}$ when buying, respectively. Note that buying
$\mathfrak{G}$ at a high price corresponds to selling money at low
price expressed in units of $\mathfrak{G}$. The squared absolute
value of the number $z$ that parameterizes the trader's strategy
\begin{equation*}
|z\rangle=|0\rangle+|z|\,\mathrm{e}^{\text{i}\arg(z)}|\text{I}\rangle
\end{equation*}
has an orthodox stochastic (non-quantum) interpretation as a
relative measure of the probability of the event (selling at high
price) with respect to the alternative event (selling at low
price) that happens with weight $1$. The phase
$\varphi\negthinspace:=\negthinspace\arg(z)$ of the parameter $z$
is characteristic of the quantum description. For
$\varphi\negthinspace=\negthinspace 0$ the absolute value of
$\frac{1-z}{1+z}$ reach its minimal value
$\frac{\left|\,1-|z|\right|}{1+|z|}\,$ and for
$\varphi\negthinspace=\negthinspace \pi$ the maximal one, equal to
the inverse of the minimal value. For
$\varphi\negthinspace=\negthinspace \frac{\pi}{2}$ and
$\varphi\negthinspace=\negthinspace \frac{3\pi}{2}$
 the absolute value of $\frac{1-z}{1+z}$ is equal
to one, what corresponds to equal probabilities $\frac{|\langle
z|0'\rangle|^2}{\langle z|z\rangle}$ and $\frac{|\langle
z|\text{I}'\rangle|^2}{\langle z|z\rangle}$\,. Changes in the
phase $\varphi$ produce no effect on the probabilities of selling
but may change the probabilities of buying corresponding to the
Fourier transformed strategies. Various price preferences of the
selling the commodity $\mathfrak{G}$ trader may interfere and
influence on his behavior as a buyer. Analogous observation can be
made about the phase of the strategy represented in the basis
$(|0'\rangle,|\text{I}'\rangle)$. (This should be compared with
behavior according to signals given by various tools used in
technical analysis.) In that way one quantum strategy may
"compete" with two independent "classical" strategies. This
mechanism enable extraordinary profits that can  hardly be
achieved in "the classical way". Quantum strategies have also
other interesting properties. For example for games that can be
represented in countably dimensional Hilbert spaces the maximal
profit is achieved at a fixed point of the tactics what gives an
effective method of adopting strategies to the continuously
changing market situation \cite{4}.
\section{Non-collective quantum tactics}
The elementary tactics of  a trader are those that result in
inverted  behavior: selling (buying) at low price is switched to
selling (buying) at high price and vice versa. Such change in the
supply strategy is described in the basis
$(|0\rangle,|\text{I}\rangle)$ by the Pauli matrix $\sigma_1$
$$
\mathcal{X}=\begin{pmatrix}
  0& 1 \\
  1 & 0
\end{pmatrix},
$$
$\mathcal{X}^2\negthinspace=\negthinspace I$. Analogously the
change in the demand strategy is described  by the Pauli matrix
$\sigma_3$
$$
\mathcal{X}'=\begin{pmatrix}
  1& \phantom{-}0 \\
  0 & -1
\end{pmatrix},
$$
because $\mathcal{X}'=\mathcal{F}\mathcal{X}\mathcal{F}$. The
matrices $\mathcal{X}$ and $\mathcal{X}'$ are interchanged by
Fourier transform therefore their sum is invariant. We have
$\mathcal{F}\negthinspace=\negthinspace\frac{1}{\sqrt{2}}\,(\mathcal{X}\negthinspace
+\negthinspace\mathcal{X}')$ in both bases. This means that for
given probabilities of switching supply to demand $\mathcal{F}$
represents a tactic. We will call any linear operation
transforming a single strategy (i.e\mbox{.} a qubit) a
non-collective tactic (a quantum gate). To illustrate the
importance of the demand-supply symmetry of the tactics
$\mathcal{X}$ i $\mathcal{X}'$ let us consider the change of the
tactics $|z\rangle$ connected with the problem of finding out if
some unknown function $g_?$ belonging to the class
$g_k\negthinspace:\{0,1\}\negthinspace\rightarrow\negthinspace\{0,1\}$,
$k\negthinspace=\negthinspace0,\ldots,3$ is constant. To this end
it is sufficient to restrict oneself to the tactics
$\mathcal{G}_?$ changing sign of the homogenous coordinates of the
strategy $|z\rangle$. In the supply basis we have:
\begin{equation*}
\mathcal{G}_k|z\rangle:=(-1)^{g_k(0)}|0\rangle+(-1)^{g_k(1)}z\,|\text{I}\rangle\,.
\label{herkiuk}
\end{equation*}
For the constant functions $g_0(0)\negthinspace=\negthinspace
g_0(1)\negthinspace=\negthinspace0$ and
$g_1(0)\negthinspace=\negthinspace
g_1(1)\negthinspace=\negthinspace1$ the respective involutions
comprise of the identities
$\mathcal{G}_0\negthinspace=\negthinspace\mathcal{G}_1\negthinspace=\negthinspace
I$ and for the remaining functions
$g_2(0)\negthinspace=\negthinspace
g_3(1)\negthinspace=\negthinspace0$ and
$g_3(0)\negthinspace=\negthinspace
g_2(1)\negthinspace=\negthinspace1$ we have the inversion of
buying strategies,
$\mathcal{G}_2\negthinspace=\negthinspace\mathcal{G}_3\negthinspace=\negthinspace
\mathcal{X}'$. Therefore we have to determine if the use of the
tactic $\mathcal{F}\mathcal{G}_k\mathcal{F}$, equal to $I$ for
$k\negthinspace=\negthinspace0,1$ and $\mathcal{X}$ for
$k\negthinspace=\negthinspace2,3$, changes the supply strategies
$|0\rangle$ or $|\text{I}\rangle$. This simple method of deciding
the question by simultaneous action of the tactics $\mathcal{G}_?$
on the superposition  of two basic supply strategies
($\mathcal{F}|0\rangle\negthinspace=\negthinspace|0'\rangle\negthinspace
=\negthinspace|0\rangle+|\text{I}\rangle$) is known as the Deutsch
Oracle \cite{9}. This simple example, although of academic
concern, shows the advantage of quantum strategies over classical
ones. One have to determine only one of the possible values of
$g_?$ to execute the strategy $\mathcal{G}_?$ because it
simultaneously acts in the parallel universes. The non-quantal
method of identification of the label $k$ of the function $g_k$ is
not so quick because it requires previous cataloguing (ie
measuring) of all
functions  $g_0$, $g_1$, $g_2$, $g_3$. \\

Any  tactic can be expressed in terms of Pauli matrices because
the second generator of the algebra $\text{su}(2)$, $\sigma_2$,
represents the tactic $\,\mathcal{X}\mathcal{X}'$ (in the basis
$(|0\rangle,|\text{I}\rangle)$). The stereographic projection
$\overline{\mathbb{C}}\negthinspace\rightarrow\negthinspace S_2$
can be readily inversed: $
S_2\negthinspace\ni(x_1,x_2,x_3)=E_z(\overrightarrow{\sigma})$,
where the vector $E_z(\overrightarrow{\sigma})= \frac{\langle
z|\overrightarrow{\sigma}|z\rangle} {\langle z|z\rangle} $
represents the expectation value of the vector of Pauli matrices
$\overrightarrow{\sigma}:=(\sigma_1,\sigma_2,\sigma_3)$ for a
given strategy $|z\rangle$. Due to this relation to the $SU(2)$
group these tactics are represented in terms of special unitary
matrices and can be parameterized by an element of $
\overline{\mathbb{C}}$ and an angle $\alpha\in[0,\pi]$:
\begin{equation} SU(2)\ni\mathcal{U}\,_{\negthinspace z,\alpha}=\,
\mathrm{e}^{\text{i}\alpha\overrightarrow{\sigma}\cdot
E_z(\overrightarrow{\sigma})}\,=\,I\cos\alpha
+\text{i}\,\overrightarrow{\sigma} \negthinspace\cdot\negthinspace
E_z(\overrightarrow{\sigma})\,\sin\alpha\,. \label{qqqutak}
\end{equation}
The coordinates
$(\cos\alpha,E_z(\overrightarrow{\sigma})\sin\alpha)
\negthinspace\in\negthinspace\mathbb{R}^4$ are in 1-1
correspondence with the sphere $S_3$
($\cos^2\alpha+E^2_z(\overrightarrow{\sigma})\,\sin^2\alpha=1$).
 The transition from the supply picture
to the demand picture is given by the homography $(\ref{herkkk})$:

\begin{equation}
\mathcal{U}_{z,\alpha} \longrightarrow
\mathcal{U}\,'_{\negthinspace z,\alpha}\negthinspace:=
\mathcal{F}\,\mathcal{U}_{z,\alpha}\,
\mathcal{F}=\mathcal{U}_{\tfrac{1-z}{1+z},\alpha}\,.
\label{gsatyr}
\end{equation}
The property $(\ref{gsatyr})$  distinguish the coordinates
$(z,\alpha)$ of a q-tactics from any other parameterization of
$S_3$. The above "market interpretation" of the group $SU(2)$
stems from Orlov approach to modelling of consciousness
\cite{10,11}. But there are important differences. The operator
$\mathcal{X}'$ describes the state of consciousness related to
subjective doubts (doubt state) concerning the truth of a given
statement  and the respective classical logic is represented by
some orthonormal basis in $\mathbb{C}^2$ (homogeneous coordinates
of a qubit) \cite{10}. Our interpretation exposes the operational
properties of $\mathcal{X}'$ (and its dual $\mathcal{X}$) that
allow the trader to completely change her demand behavior without
influencing her supplying strategy. This result in quantum game
theory is related to the construction of transactional logic in
which the values truth or false rate utility of a given strategy
measured by the respective payoffs. This utilitarian relativism is
closely related to Deutsch philosophy of science \cite{12}.
\section{Wigner function of a quantum strategy}
\begin{figure}[h]
\begin{center}
\includegraphics[height=6cm, width=9cm]{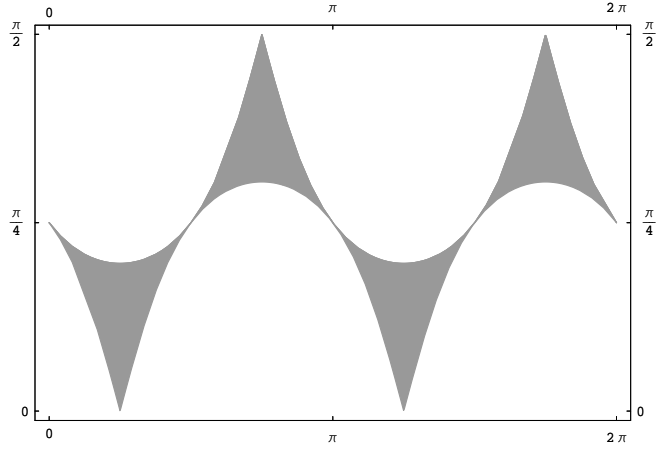}
\end{center}
\caption{Geographic map of quantum strategies. The upper curve
corresponds to two circles tangent at the north pole and are given
by the formula $\theta(\varphi)=\frac{1}{2}\,\arccos\frac{\sin
2\varphi}{2+\sin 2 \varphi}$\,.} \label{mapgif}
\end{figure}

\begin{figure}[h]
\begin{center}
\phantom{I}\vspace{-7ex}
\includegraphics[height=12cm, width=12cm]{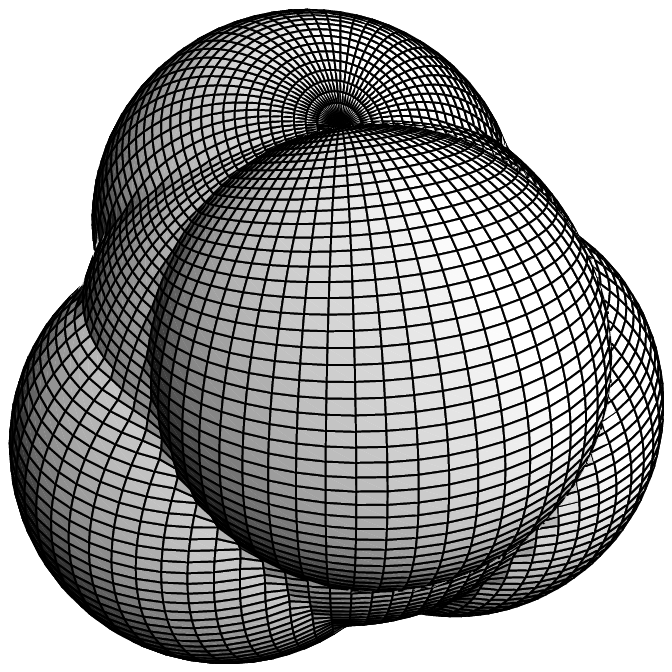}
\vspace{-16ex}
\end{center}
\caption{Parameter space ($S_2$) with radius modified by the
$\kappa$ measure of non positivity of Wigner function describing
strategies. The maximal "departure" is achieved for the strategy
$z=\pm\bigl(\frac{\sqrt{3}-1}{2}\,(1-i)\bigr)^{\pm1}$.}
\label{hklikklak}
\end{figure}

To present the effectiveness of quantum market strategies we
restrict our analysis to the most profitable situation of a game
against a "non-quantum" market with equal buying and selling
prices. Besides we will suppose that buying and selling
transaction intensities are equal. This means that the transaction
are specified by equally frequent measurements of both
polarizations $|0\rangle$ and $|0'\rangle$ \cite{2,6}. The case of
non-equal transaction intensities can be analyzed in an analogous
way by means of quantum tomography \cite{13}. If we suppose that
there are only two price levels (low and high) and the payoffs are
$1$, $0$ or $-1$ (gain -- no transaction -- loss) then the payoff
matrix takes the form given in Table \ref{hmacyp}.
\begin{table}[h]
\begin{center}
\begin{tabular}{r|c|c|c|c|}
 \multicolumn{1}{c}{} & \multicolumn{1}{c}{$\scriptstyle \hphantom{'}00'$} &
  \multicolumn{1}{c}{$\scriptstyle \hphantom{'}01'$} &
  \multicolumn{1}{c}{$\scriptstyle \hphantom{'}10'$} &
  \multicolumn{1}{c}{$\scriptstyle \hphantom{'}11'$}\\
\cline{2-5}
  {\scriptsize low price} & $-1+1=0$ & $-1+1=0$ & $0+1=1$ & $0+1=1$ \\
\cline{2-5}
  {\scriptsize high price} & $\phantom{-}1-1=0$ & $\phantom{-}1+0=1$ & $1-1=0$ & $1+0=1$ \\
\cline{2-5}
\end{tabular}
\end{center}\vspace{1ex}
\caption{The payoff matrix (trader's gains).} \label{hmacyp}
\end{table}
For example, if the price of the commodity $\mathfrak{G}$ is low
then the trader whose strategy is sell at low price ($|0\rangle$)
and buy at high price ($|0\rangle'$) loses when selling (payoff is
$-1$) and gains when buying (payoff is $1$) because being ready to
buy at high price she certainly will buy at low price. The average
payoff is
$-1\negthinspace+\negthinspace1\negthinspace=\negthinspace0$. From
the classical point of view the optimal solution is to use two
independent strategies: sell at high price  and buy at low price.
This results in average payoff $\tfrac{1}{2}$ per transaction.
What is the optimal strategy in the quantum domain?  To find out
we should determine a measure of intertwining of all possible
trader's moves. This can be done with help of the Wigner function
formalism. The Wigner function for the strategy $|z\rangle$ has
the form of a $2\negthinspace\times\negthinspace2$ matrix
\cite{14}
$$
W_{km'}(\mathrm{e}^{\text{i}\varphi}\negthinspace\tan\theta)=
\tfrac{1}{4}\bigl(1 + (-1)^k \cos 2\theta + (-1)^{m'}\sqrt{2}\sin
2\theta\cos(\varphi-(-1)^k\tfrac{\pi}{4})\bigr)\,,
$$
$k,m'=0,1$. The function $W_{km'}(z)$ gives the measure (not
always positive definite) of the state $|z\rangle$ being
simultaneously in both states $(1\negthinspace-\negthinspace
k)|0\rangle+k\,|\text{I}\rangle$ and
$(1\negthinspace-\negthinspace m')|0'\rangle+m'|\text{I'}\rangle$.
The respective  probabilities of  measurements of the strategy
being in the states $|0\rangle$, $|\text{I}\rangle$, $|0'\rangle$,
$|\text{I}'\rangle$ are :
\begin{equation*}
\begin{split}
W_{00}(z)+W_{01}(z)=\frac{|\langle z|0\rangle|^2}{\langle
z|z\rangle}\,\,&,\,\, W_{10}(z)+W_{11}(z)=\frac{|\langle
z|\text{I}\rangle|^2}{\langle
z|z\rangle}\,,\\
W_{00}(z)+W_{10}(z)=\frac{|\langle z|0'\rangle|^2}{\langle
z|z\rangle}\,\,&,\,\, W_{01}(z)+W_{11}(z)=\frac{|\langle
z|\text{I}'\rangle|^2}{\langle z|z\rangle}\,.
\end{split}
\end{equation*}
These sums being measurable in "experiment" should be nonnegative.
The same is true for the sums of diagonal elements of the matrix
$W(z)$ because they give probabilities of measurement of
strategies $|\text{i}\rangle$ and $|-\text{i}\rangle$:
\begin{equation}
W_{00}(z)+W_{11}(z)=\frac{|\langle z|\text{i}\rangle|^2}{\langle
z|z\rangle \langle \text{i}|\text{i}\rangle}\,\,,\,\,
W_{10}(z)+W_{01}(z)=\frac{|\langle z|-\text{i}\rangle|^2}{\langle
z|z\rangle\langle -\text{i}|-\text{i}\rangle}\,.
\end{equation}
This means sum of any two entries of the matrix  $W(z)$ is
nonnegative. It follows that only one of the entries of the matrix
$W(z)$ can be negative. There is a simple "quantum" strategy that
outperforms any "classical" one: one should look for such
strategies $|z\rangle$ that the negative element of the matrix
$W(z)$ corresponds to the minimal element of the payoff matrix.
Fig\mbox{.} \ref{mapgif} presents regions of the parameter space
($S_2$) that are not accessible to a "classical player". The white
regions correspond to  non positive definite matrices
$W(\mathrm{e}^{\text{i}\varphi}\tan\theta)$. Their boundaries form
four circles that correspond to the families of strategies
$z(\phi)=\pm\bigl(1+(1-\text{i})\tan\frac{\phi}{2}\bigr)$ i
$z(\phi)=\pm\frac{1}{2}(\sqrt{2}\,\mathrm{e}^{\text{i}\phi}+
1-\text{i})$, $\phi\negthinspace\in\negthinspace[-\pi,\pi)$.
If we use a measure on $S_2$ that is invariant with respect to
q-tactics then the area of the classically inaccessible region is
maximal. In that sense the quantum approach is the maximal one. It
is tempting to define a measure $\kappa$ of non-positivity  of the
Wigner function that causes the attractiveness of quantum
strategies. If we use the minimal entry of the matrix $W(z)$ to
this end we get
$\kappa(\varphi,\theta)\negthinspace:=\negthinspace
-\min\limits_{k,l} W_{kl}(\varphi,\theta)$ for regions where
$W(z)$ is not positive definite and  $\kappa(\varphi,\theta):=0$
otherwise. Fig\mbox{.} \ref{hklikklak} presents the sphere of
trader's strategies of radius $r$ modified so that this factor is
stressed (\,$r\rightarrow (1+2\,\kappa(\varphi,\theta))\,r$\,).
\section{Non-collective game against market dictating prices }
Let $p$ be the probability that the market bids a high price for
the asset $\mathfrak{G}$ and $1\negthinspace-\negthinspace p$\/ be
the probability of a low bid. If payoff matrix is given by Table
\ref{hmacyp} then the trader's average payoff is
\begin{equation*}
\begin{split}
w(p,z):=&\,E(M)=(1-p)(W_{10'}+W_{11'})+p\, (W_{01'}+W_{11'})=\\&
\,p\,(W_{01'}-W_{10'})+W_{10'}+W_{11'} =
\frac{p\,(1-z)(1-\overline{z})+2(1-p)|z|^2}{1+|z|^2}\,\,.
\end{split}
\end{equation*}
The strategy that maximizes the expected profit $w(p,z)$ is,
except for $p=0$ or $1$, a giffen \cite{2,15} that is it does not
comply with the law of demand and supply\footnote{The conditional
function of demand (supply) that corresponds to the Wigner
function formally comply with the demand and supply law but is not
measurable because it can be greater than one.}:
\begin{equation*}
z_{\max}(p)=\frac{1-p}{p}-\sqrt{1+\Bigl(\frac{1-p}{p}\Bigr)^2}.
\end{equation*}
This family of strategies comprises one fourth of  a circle
$\varphi=\pi$, $\theta\in[\frac{\pi}{4},\frac{\pi}{2}]$ in the
parameter space $S_2$. For $p=\frac{1}{2}$ we have an optimal
strategy with maximal value of $\kappa$ with Wigner matrix of the
form
\begin{equation*}
W(\pi,\tfrac{3}{8}\,\pi)=\frac{1}{4}\begin{pmatrix}
1-\sqrt{2} & 1 \\
1 & 1+\sqrt{2}
\end{pmatrix}.
\end{equation*}
\begin{figure}[h]
\begin{center}
\includegraphics[height=4.5cm, width=7cm]{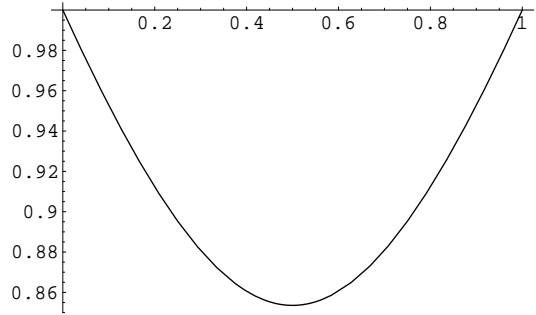}
\end{center}
\caption{Maximal profit as a function of intensity $p$ of high
price bids,
$w_{\max}=\tfrac{1}{2}+\sqrt{\tfrac{1}{8}+\tfrac{1}{2}\,
(p-\frac{1}{2})^2}$ (in a classical game maximal profit is
$\frac{1}{2}$).} \label{huwiedl}
\end{figure}

This strategy corresponds to a fixed point of the tactics
$\mathcal{XX'FX'X}$. Let us notice that in this case the worst
quantum strategy is the one that has the same form in both
representations, $z=\sqrt{2}-1$ (profit is equal to $\frac{2 -
\sqrt{2}}{4}$). Nevertheless it is better than any classical
strategy. \\

The game discussed above can be easily generalized to a realistic
case without restrictions on the price of $\mathfrak{G}$. The
resulting quantum market can be perceived as a sum of games (with
qubit strategies) for all binary digits of the logarithm of the
price of $\mathfrak{G}$. The use of logarithms makes the
considerations independent of monetary units and units used to
measure the commodity $\mathfrak{G}$.

\section{Alliances}
\begin{figure}[h]
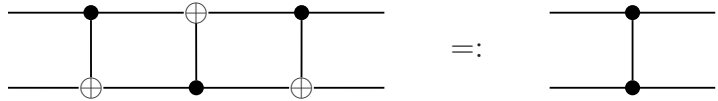

\begin{center}
\phantom{a}\vspace{5ex} \psset{linewidth=.7pt}
\rput(-4.7,1){\rnode{A}{}} \cnode*(-3.6,1){.1}{B}
\rput(-2.2,1){\rnode{C}{$\oplus$}} \cnode*(-.8,1){.1}{D}
\rput(0.3,1){\rnode{E}{}} \rput(2.5,1){\rnode{F}{}}
\cnode*(3.6,1){.1}{G} \rput(4.7,1){\rnode{H}{}}
\rput(-4.7,0){\rnode{I}{}} \rput(-3.6,0){\rnode{J}{$\oplus$}}
\cnode*(-2.2,0){.1}{K} \rput(-.8,0){\rnode{L}{$\oplus$}}
\rput(.3,0){\rnode{M}{}} \rput(2.5,0){\rnode{N}{}}
\cnode*(3.6,0){.1}{O} \rput(4.7,0){\rnode{P}{}}
\rput(1.4,.5){\rnode{Q}{$=:$}} \ncline[nodesep=0pt]{-}{A}{C}
\ncline[nodesep=0pt]{-}{C}{E} \ncline[nodesep=0pt]{-}{I}{J}
\ncline[nodesep=0pt]{-}{J}{L} \ncline[nodesep=0pt]{-}{L}{M}
\ncline[nodesep=0pt]{-}{F}{H} \ncline[nodesep=0pt]{-}{N}{P}
\ncline[nodesep=0pt]{-}{B}{J} \ncline[nodesep=0pt]{-}{C}{K}
\ncline[nodesep=0pt]{-}{D}{L} \ncline[nodesep=0pt]{-}{G}{O}
\end{center}
\caption{Tactic resulting in exchange of strategies $\mathcal{T}$
(three successive alliances).} \label{hgralyt}
\end{figure}

Let us now consider a classical market organized in such a way
that the strategies can influence each other. Suppose that there
are $N$ active players. The state of the game is represented by an
element of
$\mathbb{C}P^{\,2^N\negthinspace-\negthinspace1}$\negthinspace.
The homogeneous coordinates are formed by tensor product od the
qubits' coordinates of separate players. Such a market can be
formed for example by sufficiently large statistical ensemble of
collective strategies supplemented with an appropriate
clearinghouse that measures supply and demand and sets the
"optimal" prices. The analysis of such markets is a challenge but
the properties of unitary transformations suggest that the task
can be performed with help of quantum computation \cite{16}. In a
quantum market game \cite{2} any $N$-qubit unitary transformation
seems to be acceptable as a collective tactic because it can be
easily realized in terms of simple operations performed by
individual players or pairs of players. To this end it suffices to
define a two qubit involutive gate $\mathcal{C}$
\begin{equation*}
\mathcal{C}:=\tfrac{1}{2}\,(I+\mathcal{X}')\otimes I+
\tfrac{1}{2}\,(I-\mathcal{X}')\otimes\mathcal{X}\,
\end{equation*}
that is nontrivial only on a subspace corresponding to two
arbitrary selected players. If the first player is ready to sell
$\mathfrak{G}$ only at high price (strategy $|\text{I}\rangle$)
then the operation $\mathcal{C}$ inverses the second player's
strategy. It leaves it untouched if the first player accepts low
price when selling. We will call the gate $\mathcal{C}$ an
alliance regardless of the standard name {\em controlled-NOT}\/
(the analysis given below would explain this apostasy). If we
consider the effects of $\mathcal{C}$ in demand representation (in
the basis $(|0'\rangle),|\text{I}'\rangle)$) then the players
roles are reversed what follows from:
\begin{equation}
\mathcal{C}'=(\mathcal{F}\otimes\mathcal{F})\bigl(\tfrac{1}{2}(I+\mathcal{X}')\otimes
I+
\tfrac{1}{2}(I-\mathcal{X}')\otimes\mathcal{X}\bigr)(\mathcal{F}\otimes\mathcal{F})=
I\otimes\tfrac{1}{2}(I+\mathcal{X}')+
\mathcal{X}\otimes\tfrac{1}{2}(I-\mathcal{X}')\,. \label{hfajnal}
\end{equation}
Therefore to tell who is manipulated and who manipulates depends
on the demand/supply context.
\begin{figure}[h]
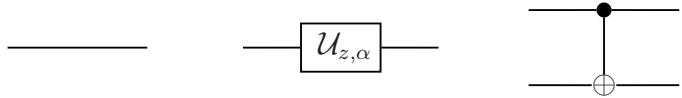

\begin{center}
\phantom{A}\vspace{3ex}
\psset{linewidth=.7pt}\mbox{\rput(-4.5,.5){\rnode{G}{}}
\rput(-2.5,.5){\rnode{H}{}} \ncline[nodesep=2pt]{-}{G}{H}
\rput(-1.3,.5){\rnode{I}{}}
\rput(0,.5){\rnode{J}{\psframebox{$\phantom{.}\mathcal{U}_{z,\alpha}$}}}
\rput(1.3,.5){\rnode{K}{}} \ncline[nodesep=0pt]{-}{I}{J}
\ncline[nodesep=0pt]{-}{J}{K} \rput(2.5,1){\rnode{A}{}}
\cnode*(3.5,1){.1}{B} \rput(4.5,1){\rnode{C}{}}
\rput(2.5,0){\rnode{D}{}} \rput(3.5,0){\rnode{E}{$\oplus$}}
\rput(4.5,0){\rnode{F}{}} \ncline[nodesep=0pt]{-}{A}{B}
\ncline[nodesep=0pt]{-}{B}{C} \ncline[nodesep=0pt]{-}{D}{E}
\ncline[nodesep=0pt]{-}{E}{F} \ncline[nodesep=0pt]{-}{B}{E}}
\end{center}
\caption{Elements of a quantum scheme (from left to right):
strategy, qubit tactic and  alliance $\mathcal{C}$.}
\label{hgrrtyyy}
\end{figure}
Most of quantum gates are universal in the sense that  any other
gate can be composed of a universal one \cite{17}. But for our
aims it is more transparent to describe a collective tactic of $N$
players as a sequence of various operations
$\mathcal{U}_{z,\alpha}$ performed on one-dimensional subspaces of
players' strategies and, possibly, alliances $\mathcal{C}$ among
them (any element of $SU(2^N)$ can be given such a form
\cite{18}). Therefore an alliance is the only way to form
collective games ($\mathcal{C}$ is universal on one qubit
subsystem). All elements necessary for implementation of such
games are presented in Figure \ref{hgrrtyyy}. Description of cases
with arbitrary number of commodities or allowed prices is obvious
but results in more complex circuits. Note that players can
exchange strategies with each other and therefore any permutation
of strategies is possible. Exchange of strategies can be
accomplished by performing three successive alliances
$\mathcal{T}\negthinspace=\mathcal{C}\,\mathcal{C}'\mathcal{C}$,
cf Fig\mbox{.} \ref{hgralyt}. Therefore Eq\mbox{.}
$(\ref{hfajnal})$ can be written in the simple form:
$\mathcal{C}'\negthinspace=\negthinspace\mathcal{T}\mathcal{C}\mathcal{T}$.
Consider a player who accepts low price for the commodity she
wants to sell. Some other player can use the alliance
$\mathcal{C}$ to change her strategy so that the resulting
correlated strategy is a two qubit entangled strategy
\begin{equation}
\mathcal{C}|z\rangle|0\rangle=\mathcal{C}|0\rangle|0\rangle+z\,\mathcal{C}|\text{I}\rangle|0\rangle
=|0\rangle|0\rangle+z\,|\text{I}\rangle|\text{I}\rangle\,.
\label{szszczegal}
\end{equation}
Strategies of selling at high prices can be
$\mathcal{C}$-transformed into anti-correlated entangled strategy
\begin{equation*}
\mathcal{C}|z\rangle|\text{I}\rangle=|0\rangle|\text{I}\rangle+z\,|\text{I}\rangle|0\rangle\,.
\end{equation*}
Analogous manipulations (transposition) are possible in the demand
part of the market.
\section{Alliances: measurement and interference}
\begin{figure}[h]
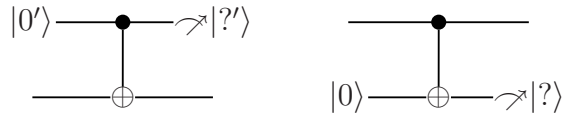

\begin{center}
\phantom{a}\vspace{5ex} \psset{linewidth=.7pt}
\rput(-3.3,1){\rnode{A}{$|0'\rangle$\hspace{1pt}}}
\cnode*(-2.1,1){.1}{B}
\rput(-0.9,1){\rnode{C}{\meter$|?'\rangle$}}
\rput(0.9,1){\rnode{D}{}} \cnode*(2.1,1){.1}{E}
\rput(3.3,1){\rnode{F}{}} \rput(-3.3,0){\rnode{G}{}}
\rput(-2.1,0){\rnode{H}{$\oplus$}} \rput(-0.9,0){\rnode{I}{}}
\rput(0.9,0){\rnode{J}{$|0\rangle$\hspace{1pt}}}
\rput(2.1,0){\rnode{K}{$\oplus$}}
\rput(3.3,0){\rnode{L}{\meter$|?\rangle$}}
\ncline[nodesep=0pt]{-}{A}{C} \ncline[nodesep=0pt]{-}{D}{F}
\ncline[nodesep=0pt]{-}{G}{H} \ncline[nodesep=0pt]{-}{H}{I}
\ncline[nodesep=0pt]{-}{J}{K} \ncline[nodesep=0pt]{-}{K}{L}
\ncline[nodesep=0pt]{-}{B}{H} \ncline[nodesep=0pt]{-}{E}{K}
\end{center}
\caption{Alliance as a means of determining others' strategies.
The sign ,,\meter\hspace{.06em}'' at the right ends of lines
representing qubits symbolizes measurement.} \label{sulokitek}
\end{figure}

An alliance allows the player to determine the supply or demand
state of another player by making an alliance and measuring her
resulting strategy. This process is shortly described as
\begin{equation*}
\mathcal{C}\,|0'\rangle|m'\rangle=|m'\rangle|m'\rangle,\,\,\,\,
\mathcal{C}\,|m\rangle|0\rangle=|m\rangle|m\rangle,
\end{equation*}
where $m\negthinspace=\negthinspace0,\text{I}$. The corresponding
diagrams are shown in Fig\mbox{.} \ref{sulokitek}. The left diagram
presents measurement of the observable $\mathcal{X}$ and the right
one  measurement of $\mathcal{X}'$.
Any measurement would demolish possible entanglement of
strategies.  Therefore entangled quantum strategies can exist only
if the players in question  are ignorant of the details of their
strategies. Let us now consider an interesting version of the game
when the players have "quantum minds" (or are sort of
self-conscious quantum automata \cite{19}). According to the many
worlds interpretation of quantum theory \cite{12} such a player is
aware of her strategy belonging to any of the interfering worlds.
If the worlds do not decouple for some time and strategies are
transformed according to the applied tactics and if eventually the
measurement is performed then such a player would not be aware of
the decoupled histories. Any quantum mind that wants to evolve in
a unitary (deterministic) quantum way is involuntary as being
condemned to a quantum evolution (cf \cite{20} for a discussion of
the free choice problem). Such a quantum mind may try to influence
his evolution by a tactics profitable from the point of view of
that component of the interfering strategies that she is aware of.
But such  operation announce information about her (quantum) state
and is equivalent to a measurement that would destroy any
extraordinary profits resulting from interference with other
worlds. However, the possibility of choosing between destroying
and preserving the interference alone gives her certain advantages
over the classical minds. Only passive persistence in the
interference due to the necessary amnesia would postpone  the
eventual valuation of her behavior. Details of conclusions one can
draw may depend on the actual interpretation of quantum theory one
is ready to accept. Nevertheless, parties playing a quantum game
must be very careful  in controlling their strategies and the very
measurement should form an inseparable ingredient of
tactic/strategy. Therefore the limited knowledge of players'
strategy result in a characteristic for a quantum game
spontaneousness. Eq\mbox{.} $(\ref{qqqutak})$ and the identities:
\begin{equation}
\begin{split}
(I\otimes\mathcal{X}
)\,\mathcal{C}&=\mathcal{C}\,(I\otimes\mathcal{X}),\\
(\mathcal{X}'\otimes I
)\,\mathcal{C}&=\mathcal{C}\,(\mathcal{X}'\otimes I),\\
(\mathcal{X}\otimes
I)\,\mathcal{C}&=\mathcal{C}\,(\mathcal{X}\otimes\mathcal{X}),\\
(I\otimes\mathcal{X}'
)\,\mathcal{C}&=\mathcal{C}\,(\mathcal{X}'\otimes\mathcal{X}')
\end{split}
\label{kuuudela}
\end{equation}
imply that any game (collective strategy) can be perceived as
superposition of some simpler collective strategies that result
from two stages (one can reverse the time arrow). First, players
form alliances and then some of them inverse supply $\mathcal{X}$,
$\mathcal{X}'$ or both $\mathcal{XX}'$. Transposition of alliances
and qubit inversions does not change anything or results in
cloning of qubits. \\

The following digressions seems to be in place here. In 1970
Sthephen Wiesner invented counterfeit-proof (quantum) money
\cite{8}. The idea was based on the first proposal of a one-way
function that exploits unique properties of quantum states. This
proposal contrary to its mathematical counterparts \cite{21}
cannot be questioned because fundamental laws of Nature guarantee
its properties. Wiesner's proposal was put forward three years
earlier than Cock's idea of using  difficult to reverse operations
in cryptography (cf the construction of safe quantum cryptography
by Bennett and Brassard in 1984 that cannot be broken even with
the help of quantum computers) \cite{22}. According to Wiesner a
counterfeit-proof banknote should have two numbers, one of them is
kept secret. The secret number is formed from two random binary
sequences and is encoded  in a two-state quantum systems as one of
the four states $|0\rangle,|\text{I}\rangle, |0'\rangle$ and $
|\text{I}'\rangle$ in accordance with a scheme presented in
\cite{8}. The second number plays the role of an ordinary serial
number. The knowledge of the two randomly generated binary
sequences allows to perform a non-destroying verification
(measurement) of the authenticity of the note (the proper method
of reading of the qubits forms the trap-door of Wiesner's one-way
function). The demand or supply formal aspect of the binary digit
requires a correct method of measurement (left of right part in
Fig\mbox{.} \ref{sulokitek}). Wiesner's pioneering idea of quantum
banknote and Deutsch Oracle show that the mathematical notion of a
function might be more realistic than one is ready to admit.
Present development of quantum information theory is a case in
point for creativity of such attitude. Properties of quantum world
may seem to be strange but  always comply with objective laws of
Nature.

\section{Some remarks on collective market games}
A game when only "classical" tactics $\mathcal{T}$ are allowed can
be used for lottery-drawing when there is no drawing machine
available \cite{23}. Such a drawing can have the following form.
The person that carries the drawing out (she might also take part
in the drawing) draws parallel horizontal lines on a sheet of
paper that correspond to the participants. Then she marks the left
end of one the lines bends the sheet so that the other participant
cannot see the left ends. At that moment the proper drawing
begins: every participant draws arbitrary number of vertical lines
\begin{figure}[h]
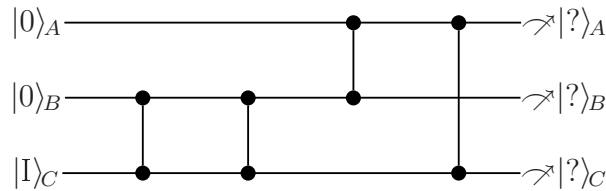

\begin{center}
\phantom{a}\vspace{10ex} \psset{linewidth=.7pt}
\rput(-3.5,2){\rnode{A}{$|0\rangle_{\negthinspace
A}$\hspace{1pt}}}
\rput(-3.5,1){\rnode{B}{$|0\rangle_{\negthinspace
B}$\hspace{1pt}}}
\rput(-3.5,0){\rnode{C}{$|\text{I}\rangle_{\negthinspace
C}$\hspace{1pt}}} \cnode*(-2.1,0){.1}{E} \cnode*(-2.1,1){.1}{D}
\cnode*(-.7,0){.1}{G} \cnode*(-.7,1){.1}{F} \cnode*(.7,1){.1}{I}
\cnode*(.7,2){.1}{H} \cnode*(2.1,2){.1}{J} \cnode*(2.1,0){.1}{K}
\rput(3.5,0){\rnode{L}{\meter$|?\rangle_{\negthinspace C}$}}
\rput(3.5,1){\rnode{M}{\meter$|?\rangle_{\negthinspace B}$}}
\rput(3.5,2){\rnode{N}{\meter$|?\rangle_{\negthinspace A}$}}
\ncline[nodesep=0pt]{-}{A}{N} \ncline[nodesep=0pt]{-}{B}{M}
\ncline[nodesep=0pt]{-}{C}{L} \ncline[nodesep=0pt]{-}{D}{E}
\ncline[nodesep=0pt]{-}{F}{G} \ncline[nodesep=0pt]{-}{H}{I}
\ncline[nodesep=0pt]{-}{J}{K}
\end{center}
\caption{Drawing without drawing machine --- the system is built
from transpositions $\mathcal{T}$ alone.} \label{ohgralosko}
\end{figure}
that join two horizontal lines. The winner is that person whose
line's end shows the way out of the maze built up according to the
drawing (one starts at the marked point, goes to the right and
must go along each vertical line met; going back or turning to the
left is forbidden). For example Alice (A) win in the situation
presented in Fig\mbox{.} \ref{ohgralosko}. The game is fair
because the result is given by a permutation composed from random
transpositions. \\

Let us now consider a collective game that has no classical
counterpart. Let as call it {\it Master and pupil}\/. Alice (A) is
ready to sell the asset $\mathfrak{G}$ at low price and Bob (B)
wants to buy $\mathfrak{G}$ even at high price. But Bob, instead
of making the deal (according to the measured strategies), enters
into an alliance with Alice. Aftermath Alice changes her strategy
according to the tactic $(\ref{qqqutak})$ and end enters into an
alliance with Bob.
\begin{figure}[h]
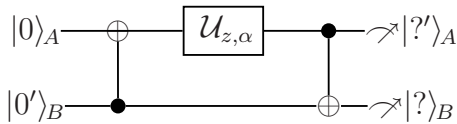

\begin{center}
\phantom{a}\vspace{5ex} \psset{linewidth=.7pt}
\rput(-2.2,1){\rnode{A}{$|0\rangle_{\negthinspace A}$}}
\rput(-1.1,1){\rnode{B}{$\oplus$}}
\rput(.3,1){\rnode{C}{\psframebox{$\phantom{.}\mathcal{U}_{z,\alpha}$}}}
\cnode*(1.7,1){.1}{D}
\rput(2.8,1){\rnode{E}{\meter$|?'\rangle_{\negthinspace A}$}}
\rput(-2.2,0){\rnode{I}{$|0'\rangle_{\negthinspace B}$}}
\cnode*(-1.1,0){.1}{J} \rput(1.7,0){\rnode{L}{$\oplus$}}
\rput(2.8,0){\rnode{M}{\meter$|?\rangle_{\negthinspace B}$}}
\ncline[nodesep=0pt]{-}{A}{C} \ncline[nodesep=0pt]{-}{C}{E}
\ncline[nodesep=0pt]{-}{I}{J} \ncline[nodesep=0pt]{-}{J}{L}
\ncline[nodesep=0pt]{-}{L}{M} \ncline[nodesep=0pt]{-}{B}{J}
\ncline[nodesep=0pt]{-}{D}{L}
\end{center}
\caption{The game {\it Master and pupil}\/ (dense coding\index{dense
coding}).} \label{hgralytyt}
\end{figure}
As a result an entangled quantum state
$|z,\alpha\rangle_{\negthinspace
AB}\negthinspace\in\mathbb{R}P^{3}
\negthinspace\subset\mathbb{C}P^{3}$ is formed, cf Fig\mbox{.}
\ref{hgralytyt}:
\begin{equation}
|z,\alpha\rangle_{\negthinspace
AB}:=\mathcal{C}\,(\mathcal{U}_{z,\alpha}\negthinspace \otimes
I)\,\mathcal{C}'\,|0\rangle_{\negthinspace
A}|0'\rangle_{\negthinspace B}= \label{ssplanqa}
\end{equation}
\begin{equation*}
\cos(\alpha)\,|0'\rangle_{\negthinspace A}|0\rangle_{\negthinspace
B} +\,\text{i}\,\sin(\alpha)\,\bigl(E_z(\mathcal{X})\,
|0'\rangle_{\negthinspace A}|\text{I}\rangle_{\negthinspace B}
+E_z(\mathcal{X}')\,|\text{I}'\rangle_{\negthinspace
A}|0\rangle_{\negthinspace B}+
E_z(\mathcal{X}\mathcal{X}')\,|\text{I}' \rangle_{\negthinspace
A}|\text{I}\rangle_{\negthinspace B}\bigr)\,.\vspace{1ex}
\end{equation*}
Although Bob cannot imitate Alice tactic $\mathcal{U}_{z,\alpha}$
by simple cloning of the state, he can gather substantial
knowledge about her strategy when she is buying (he is able to
measure proportions among the components $I$, $\mathcal{X}$,
$\mathcal{X}'$ and $\mathcal{X}\mathcal{X}'$). The game is
interesting also from Alice point of view because it allows her to
form convenient correlations of her strategy with Bob's. Such
procedure is called dense coding in quantum information theory
\cite{24}. If Alice and Bob are separated from each other and have
formed the entangled state $|0\rangle_A|0\rangle_B\negthinspace+
|\text{I}\rangle_A|\text{I}\rangle_B$ (this is the collective
strategy before the execution of
$\mathcal{U}_{z,\alpha}\negthinspace\otimes I$\/) then Alice is
able to communicate her choice of tactic ($I$, $\mathcal{X}$,
$\mathcal{X}'$\negthinspace, $\mathcal{X}\mathcal{X}'$) to Bob
(bits of information) by sending to him a single qubit. Bob can
perform a joint measurement of his and Alice's qubits.  Only one
of four orthogonal projections on the states
$|0'\rangle_{\negthinspace A}|0\rangle_{\negthinspace B}$,
$|0'\rangle_{\negthinspace A}|\text{I}\rangle_{\negthinspace B}$,
$|\text{I}'\rangle_{\negthinspace A}|0\rangle_{\negthinspace B}$
and $|\text{I}'\rangle_{\negthinspace
A}|\text{I}\rangle_{\negthinspace B}$ will give a positive result
forming the message\footnote{Answers to the questions {\it Would
Alice buy at high price?} and {\it Would Bob sell at low price?}
would decode the message.}. Such concise communication is
impossible for classical communication channels and any attempt at
eavesdropping would irreversibly destroy the quantum coherence
(and would be detected).
\begin{figure}[h]
\begin{center}
\phantom{a}\vspace{10ex} \psset{linewidth=.7pt}
\rput(-2.1,2){\rnode{A}{$|z\rangle$\hspace{1pt}}}
\rput(-2.1,1){\rnode{B}{$|0'\rangle$\hspace{1pt}}}
\rput(-2.1,0){\rnode{C}{$|0\rangle$}}
\rput(-.7,0){\rnode{E}{$\oplus$}} \cnode*(-.7,1){.1}{D}
\cnode*(.7,2){.1}{G} \rput(.7,1){\rnode{F}{$\oplus$}}
\rput(.7,0){\rnode{U}{\psframebox{$\phantom{.}\mathcal{U}_{m'n}$}}}
\rput(2.1,1){\rnode{I}{\meter $|n\rangle$}}
\rput(2.1,2){\rnode{H}{\meter $|m'\rangle$}}
\rput(2.1,0){\rnode{L}{$\,|z\rangle$}}
\ncline[nodesep=0pt]{-}{A}{H} \ncline[nodesep=0pt]{-}{B}{I}
\ncline[nodesep=1pt]{-}{C}{U} \ncline[nodesep=1pt]{-}{U}{L}
\ncline[nodesep=0pt]{-}{D}{E} \ncline[nodesep=0pt]{-}{F}{G}
\end{center}
\caption{Teleportation of the strategy $|z\rangle$ consisting in
measurement of the tactic
$\mathcal{U}_{m'n}\negthinspace:=\mathcal{X}^{[
n=\text{I}]}\mathcal{X}'^{[m'\negthinspace=\text{I}']}$ (the
notation $[true]\negthinspace:=\negthinspace1$ and
$[false]\negthinspace:=\negthinspace0$ is used).} \label{obgrewa}
\end{figure}

If one player forms an alliance with another that has already
formed another alliance with a third player then the later can
actually perform measurements that will allow him to transform his
strategy to a strategy that is identical to the first player's
primary strategy (teleportation \cite{25, 26}).  This is possible
due to the identity (remember that $\mathcal{X}$,
$\mathcal{X}'\negthinspace$, $\mathcal{X}\mathcal{X}'$ are
involutive maps)
\begin{equation*}
2\,(\mathcal{C}\otimes I)\,(I\otimes
\mathcal{C})\,|z\rangle|0'\rangle|0\rangle=
|0'\rangle|0\rangle|z\rangle+|0'\rangle|\text{I}\rangle\mathcal{X}|z\rangle
+|\text{I}'\rangle|0\rangle\mathcal{X}'|z\rangle+
|\text{I}'\rangle|\text{I}\rangle\mathcal{X}\mathcal{X}'|z\rangle\,.
\end{equation*}
Recall that quantum strategies cannot be clonned (no-cloning
theorem) but if there are several identical strategies their
number cannot be reduced by classical means (no-reducing theorem).
\\
Obviously, the most effective way of playing collective quantum
market game is as follows. One have adopt such tactic that
transform the initial strategy to one giving maximal profit,
$|z_{\max}(p)\rangle$. It seems that the players alone should
decide what forms of tactics are allowed in fact they should
construct the whole market. For example, players are allowed to
tackle only their own qubits, form a limited number of alliances
that can be secret or public. Such an autonomous market should
have sort of clearinghouse that besides measuring the respective
tactics charges some brokerage. \\
Collective quantum market games can used for convenient allocation
of rules of disposition of the asset $\mathfrak{G}$.  In that case
the direct aim of a player is to get an appropriate quantum
entanglement instead of maximization of profit. Sort of envy-free
rules can also be introduced. For example, an alliance formed at
later stages of the game gives substantial advantage. Therefore
players that have already gave up forming alliances may be allowed
to exchange strategies ($\mathcal{C}\mathcal{C}'\mathcal{C}$)
according to the Banach-Knaster protocol \cite{27}. For
$N\negthinspace=\negthinspace2$ this protocol is used in fair
division: {\it one divides and the other
chooses}\footnote{Mediterranean civilizations have been using this
method for at least 2800 years \cite{27}. It has been used in
popular games in the Far East (http://www.playsite.com). Nowadays
it is used in the game of {\it hex} invented by Piet Hein in 1942
and reinvented by John Nash. V. Anshelevich  has put forward and
effective algorithm for playing {\it hex} that is equivalent to
finding maximal current in Kirchoff's circuit \cite{28} and
probably has
interesting connections with quantum games.}. \\

The knowledge about the topology of formed alliances influences
the actual method of looking for  optimal tactics and their form.
The investigation of effective methods of playing collective
quantum market games is very complicated even in simplified
version therefore it will be postponed to a subsequent work.
\section{Conclusions}
Quantum game theory \cite{9},\cite{29}-\cite{31} could have not
emerged earlier because a tournament quantum computer versus
classical one is not possible without technological development
necessary for a construction of quantum computers. Traders active
on the markets of future would not hesitate to take the advantage
of technological development. The analysis  presented here shows
that quantum market games or more general quantum-like approach to
market description might turn out to be an important theoretical
tool for investigation of computability problems in economics or
game theory even  if never implemented in real markets.


\end{document}